\documentclass[preprint,showkeys,superscriptaddress,amsmath,amssymb]{revtex4-1}

\usepackage{newtxtext,newtxmath}
\usepackage[T1]{fontenc}
\usepackage{graphicx}
\usepackage{bm}
\usepackage{textcomp}
\usepackage{braket}
\usepackage{siunitx}
\usepackage{enumitem}
\usepackage{caption}

\DeclareCaptionLabelSeparator{bar}{ $|$ }
\usepackage[colorlinks=true,
linkcolor=magenta,
citecolor=blue,
pdfauthor={Hyounghan Kwon},
]{hyperref}

\usepackage{times}
\usepackage{dcolumn}
\usepackage{amsmath}
\usepackage[dvipsnames]{xcolor}
\usepackage[labelsep=space]{caption}
\usepackage{setspace}
\usepackage{amssymb}
\usepackage{amsmath}
\usepackage[version=3]{mhchem}
\usepackage{lineno}

\newcommand{\NatureFormat}{%
		\renewcommand{\figurename}{\textbf{Fig.}}
     }
\NatureFormat

\begin{document}

\title{Integrated Bright Source of Polarization-Entangled Photons Using Lithium Niobate Photonic Chips}

\author{Changhyun Kim}
\thanks{These authors contributed equally}
\affiliation{Center for Quantum Technology, Korea Institute of Science and Technology (KIST), Seoul 02792, South Korea}

\author{Hansol Kim}
\thanks{These authors contributed equally}
\affiliation{Center for Quantum Technology, Korea Institute of Science and Technology (KIST), Seoul 02792, South Korea}

\author{Minho Choi}
\affiliation{Center for Quantum Technology, Korea Institute of Science and Technology (KIST), Seoul 02792, South Korea}
\affiliation{Department of Artificial Intelligence Semiconductor Engineering, Hanyang University, Seoul, 04763, South Korea}

\author{Junhyung Lee}
\affiliation{Center for Quantum Technology, Korea Institute of Science and Technology (KIST), Seoul 02792, South Korea}
\affiliation{School of Electrical Engineering, Korea University, Seoul 02841, South Korea}

\author{Yongchan Park}
\affiliation{Center for Quantum Technology, Korea Institute of Science and Technology (KIST), Seoul 02792, South Korea}
\affiliation{School of Electrical Engineering, Korea University, Seoul 02841, South Korea}

\author{Sunghyun Moon}
\affiliation{Center for Quantum Technology, Korea Institute of Science and Technology (KIST), Seoul 02792, South Korea}

\author{Jinil Lee}
\affiliation{Center for Quantum Technology, Korea Institute of Science and Technology (KIST), Seoul 02792, South Korea}
\affiliation{Division of Nano and Information Technology, KIST School, Korea University of Science and Technology, Seoul, 02792, Korea}

\author{Hyeon Hwang}
\affiliation{Department of Physics, Korea Advanced Institute of Science and Technology (KAIST), Daejeon, 34141, South Korea}

\author{Min-Kyo Seo}
\affiliation{Department of Physics, Korea Advanced Institute of Science and Technology (KAIST), Daejeon, 34141, South Korea}

\author{Yoon-Ho Kim}
\affiliation{Department of Physics, Pohang University of Science and Technology (POSTECH), Pohang, 37673, Korea}

\author{Yong-Su Kim}
\affiliation{Center for Quantum Technology, Korea Institute of Science and Technology (KIST), Seoul 02792, South Korea}
\affiliation{Division of Quantum Information, KIST School, Korea University of Science and Technology, Seoul 02792, South Korea}

\author{Hojoong Jung}
\email{Co-Corresponding author: H.J.: hojoong.jung@kist.re.kr}
\affiliation{Center for Quantum Technology, Korea Institute of Science and Technology (KIST), Seoul 02792, South Korea}

\author{Hyounghan Kwon}
\email{Co-Corresponding author: H.K.: hyounghankwon@kist.re.kr}
\affiliation{Center for Quantum Technology, Korea Institute of Science and Technology (KIST), Seoul 02792, South Korea}
\affiliation{Division of Quantum Information, KIST School, Korea University of Science and Technology, Seoul 02792, South Korea}

\maketitle

\textbf{Abstract.}  
Quantum photonics has rapidly advanced as a key area for developing quantum technologies by harnessing photons' inherent quantum characteristics, particularly entanglement. Generation of entangled photon pairs, known as Bell states, is crucial for quantum communications, precision sensing, and quantum computing. While bulk quantum optical setups have provided foundational progress, integrated quantum photonic platforms now offer superior scalability, efficiency, and integrative potential. In this study, we demonstrate a compact and bright source of polarization-entangled Bell state utilizing continuous-wave pumping on thin film lithium niobate (TFLN) integrated photonics. Our periodically poled lithium niobate device achieves on-chip brightness of photon pair generation rate of 508.5 MHz/mW, surpassing other integrated platforms including silicon photonics. This demonstration marks the first realization of polarization entanglement on TFLN platforms. Experimentally measured metrics confirm high-quality entangled photon pairs with a purity of 0.901, a concurrence of 0.9, and a fidelity of 0.944. We expect our compact quantum devices to have great potential for advancing quantum communication systems and photonic quantum technologies.




\section*{Introduction}

Quantum photonics has emerged as a pivotal platform for the advancement of quantum science and technologies, leveraging the fundamental quantum characteristics of photons such as superposition and entanglement~\cite{o2007optical,kok2010introduction, walmsley2015quantum, pelucchi2022potential}. A central task in quantum photonics is the controlled generation, modulation, and detection of multi-photon quantum states, particularly maximally entangled two photon states which are commonly referred to as Bell states~\cite{anwar2021entangled}. These states are typically characterized through quantum state tomography and verified by the violation of Bell's inequality~\cite{clauser1969proposed}.These entangled states have become the cornerstone for various quantum information science including entanglement distribution~\cite{yin2012quantum, yin2017satellite}, quantum teleportation~\cite{bouwmeester1997experimental}, entanglement swapping~\cite{pan1998experimental}, entanglement purification~\cite{pan2001entanglement}, and generation of multi-photon entangled states~\cite{lu2007experimental}. Finally, large-scale photonic entangled states, often constructed from multiple Bell states, are essential for diverse quantum information protocols and systems, including quantum cryptography~\cite{ekert1991quantum}, quantum networks~\cite{kimble2008quantum}, quantum sensing~\cite{liu2021distributed, kim2024distributed}, and measurement-based quantum computing~\cite{knill2001scheme, bartolucci2023fusion}. While the two-photon Bell states have been extensively demonstrated across various degrees of freedom of photon, polarization entanglement offers particular advantages in terms of simplicity in implementation, ease of manipulation, and compatibility with standard polarization optical components. Indeed, many of the seminal quantum experiments have relied on polarization-entangled Bell states for their implementation~\cite{yin2012quantum, yin2017satellite, bouwmeester1997experimental, pan1998experimental, pan2001entanglement,lu2007experimental,kim2024distributed}.

Significant advancements in Bell state generation have been driven by bulk and fiber optics for decades. Experimental systems mostly utilize either spontaneous parametric down-conversion (SPDC), which is based on second-order nonlinear crystal~\cite{anwar2021entangled}, or spontaneous four-wave mixing (SFWM), which relies on third order nonlinearity~\cite{chen2005two}. Upon these successful foundations, integrated photonics have emerged as a promising avenue for further progress in quantum photonics~\cite{wang2018multidimensional, pelucchi2022potential, luo2023recent, bao2023very, wang2024progress, chen2021quantum}. The integrated quantum photonic devices simultaneously offer substantial advantages such as scalability, compact footprint, high stability, increased nonlinear interaction efficiency, and programmability. These attributes of the integrated photonic devices are expected to facilitate transitions from the conventional optical system to miniaturized and scalable system for broader implementation of quantum optical systems. Specifically, silicon or silicon-nitride quantum photonic devices have enabled diverse types of Bell state generation, utilizing rich degrees of freedom of photon such as time-bin, polarization, frequency-bin, and spatial-modes~\cite{chen2021quantum}. In particular, several schemes for compact polarization Bell-state generation have been experimentally demonstrated~\cite{matsuda2012monolithically, wen2023polarization, zhang2024polarization, miloshevsky2024cmos}. Nevertheless, the silicon-based quantum photonic chips may exhibit lower brightness compared to SPDC sources, and face stringent requirements for pump filtering as well as challenges in degenerate photon pair generation~\cite{moody2020chip}.

Among various material platforms in the fields of integrated photonics, thin-film lithium niobate (TFLN) has recently been considered one of the most promising platforms for quantum photonics. That is because TFLN uniquely exhibits strong second-order nonlinearity, a wide transparency window, electro-optic tunability, and compatibility with standard nanofabrication techniques~\cite{zhu2021integrated}. As a result, TFLN-based integrated photonics have enabled the efficient generation and manipulation of nonlinear quantum photonic states~\cite{jankowski2021dispersion}. For example, efficient photon pair generation~\cite{chen2019efficient, elkus2019generation, zhao2020high, xin2022spectrally, fang2024efficient, kwon2024photon, hwang2024spontaneous, kim2025freeform}, compact vacuum squeezers~\cite{lenzini2018integrated, chen2022ultra, peace2022picosecond, nehra2022few, park2024single, arge2024demonstration}, and active quantum state manipulations~\cite{zhu2022spectral, chapman2024chip} have been demonstrated. Although integrated TFLN quantum photonic devices have enabled high-performance nonlinear quantum light sources, the experimental demonstration of entangled states is still limited. Very recently, time-bin entangled photon pair sources have been demonstrated using a single periodically poled lithium niobate (PPLN) waveguide, an unbalanced Mach–Zehnder interferometer with a on-chip delay line, and a pulsed light source~\cite{finco2024time}. Despite such great advancements of TFLN quantum photonics, the generation source of polarization Bell states on TFLN chip has remained elusive. Therefore, the successful on-chip generation of polarization Bell states using TFLN could overcome the limitations of silicon-based platforms and at the same time provide a fundamental and versatile quantum light source for scalable quantum photonic applications.

In this paper, we develop a compact bright source of polarization-entangled photons leveraging thin-film lithium-niobate integrated photonics. Our design utilizes a combination of a multi-mode interferometer (MMI), two independent PPLN waveguides, and a polarization splitter-rotator (PSR) to efficiently generate polarization-entangled photon pairs. Leveraging continuous-wave (CW) pumping, our PPLN device shows high on-chip SPDC brightness of 508.5 MHz/mW, surpassing other integrated platforms, including silicon photonics. Furthermore, we perform quantum state tomography of the polarization-entangled state. Purity of 0.901, concurrence of 0.900, and fidelity of 0.944 are experimentally observed, directly indicating that high-quality polarization-entangled photon sources are generated. This integrated architecture enables compact Bell state generation, making it suitable for various applications in quantum photonics.

\clearpage

\section*{Results}
\subsection*{Concept of on-chip source of polarization-entangled state}
\begin{figure}
    \includegraphics[width=1\linewidth]{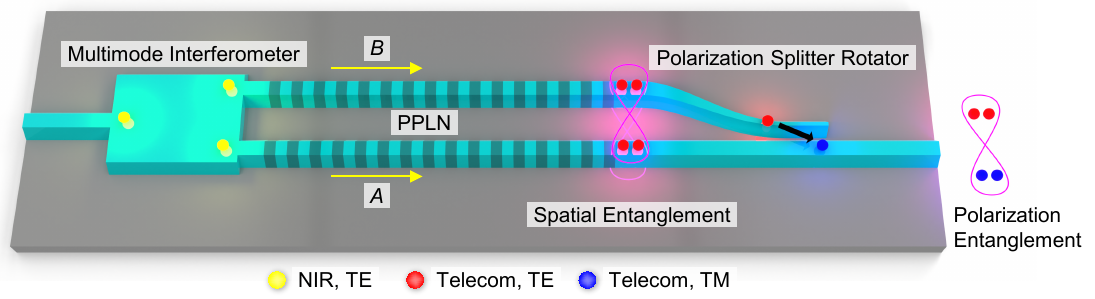}
    \caption{\label{fig:fig1} \textbf{Conceptual schematic illustration of on-chip polarization entangled photon pair generation.} Near infrared (NIR) pump is split into two path: $A$ and $B$. At each path, spontaneous parametric down conversion occurs at periodically poled lithium niobate, making spatial mode entangled state. A polarization splitter rotator convert the spatial entangled state to the polarization entangled state.}
\end{figure}

To generate the polarization-entangled photon pairs, we design an integrated photonic device on X-cut TFLN, as shown in Figure~\ref{fig:fig1}. Our device incorporates several key components, including a MMI, two PPLN waveguides, and a PSR. The generation process begins by coupling a near-infrared pump laser, which is described by a coherent state $\left|\alpha\right\rangle$, into the $1\times2$ MMI through its single input port. Assuming symmetry, the output state from the MMI, $\left|\psi_{\mathrm{MMI}}\right\rangle$, is described as a tensor product of the two coherent states split evenly, as expressed by Equation~\ref{eq:mmi}.

\begin{equation}
    \left|\psi_{\mathrm{MMI}}\right\rangle = \left|\frac{1}{\sqrt{2}}\alpha\right\rangle_A \otimes \left|\frac{1}{\sqrt{2}}\alpha\right\rangle_B
    \label{eq:mmi}
\end{equation}
where $A$ and $B$ represents the two pathways after passing through the $1\times2$ MMI device.

In each pathway, the pump undergoes type-0 SPDC in the two PPLN waveguides. In type-0 SPDC, the horizontally polarized (transverse electric, TE) coherent pump state generates photon pairs of the same horizontal polarization via the down-conversion of the pump photons from the vacuum state~\cite{silverstone2014chip}. After discarding the residual pump photons, the photon pairs in each pathway become spatially entangled, as described by Equation \ref{eq:ppln}.

\begin{equation}
    \left|\psi_{\mathrm{PPLN}}\right\rangle = \sqrt{\eta_{l}} \left|HH\right\rangle_A + \sqrt{1-\eta_{l}}\left|HH\right\rangle_B
    \label{eq:ppln}
\end{equation}
where the $\eta_{l}$ represents combined normalization factor, accounting the nonlinear interaction efficiency of the PPLN waveguides. The subscripts $A$ and $B$ denote the respective spatial pathways. In this stage, the photon pairs are in spatial-mode entanglement. Ideally, if two PPLN waveguides are made identically, the nonlinear interaction efficiencies of both will be identical as $\eta_{l} = 0.5$. However, in practical, there would be some differences in the ratio of overall efficiency due to fabrication imperfection.

To convert the spatial-mode entanglement into polarization entanglement, we utilize a polarization splitter rotator (PSR). The PSR plays a crucial role by introducing polarization rotation of TE mode in cross port (spatial mode $B$) to transverse magnetic (TM) mode, while maintaining the original TE polarization state in through port (spatial mode $A$). Since both photons occupy the same spatial mode, they must be separated for further processing. This requires probabilistic photon separation method, such as beam splitters in free-space setups, or MMI and directional couplers in integrated photonic platform. Consequently, the final generated state is represented by Equation~\ref{eq:output_state}.

\begin{equation}
     \left|\psi\right\rangle = \sqrt{\eta} \left|HH\right\rangle + \sqrt{1-\eta}e^{i2\phi} \left|VV\right\rangle
    \label{eq:output_state}
\end{equation}
where $\eta$ is the normalized efficiency ratio, and $\phi$ denotes the local phase change associated with the transition from spatial mode $B$ to spatial mode $A$ with rotating polarization, induced by PSR and modal phase dispersion. As shown in the Equation~\ref{eq:output_state}, the output state cannot be separated with their polarization state, indicating that two photons are entangled in their polarization. A more detailed quantum optical description of these dynamics can be found in Supplementary Notes 1. Also, the fabricated devices are displayed in Supplementary Figure S1.

\clearpage

\subsection*{Design and validation of building blocks}
We first design and investigate the fundamental building blocks including PPLN waveguide and PSR. For the PPLN waveguide, the height and etch depth are set to 300 nm and 240 nm, respectively, in accordance with the PSR design parameters. To determine the poling period required for achieving the quasi-phase matching (QPM) condition, we perform numerical simulations to compute the effective refractive index using commercially available finite element method, COMSOL Multiphysics. The poling period, $\Lambda$, can be calculated with the effective refractive index, $n$, as:

\begin{equation}
    \Lambda_{\mathrm{PP}} = \left( \frac{\lambda_{\mathrm{SH}}}{n_{\mathrm{SH}}-n_{\mathrm{FH}}} \right)
    \label{eq:poling_period}
\end{equation}

\begin{figure}
    \includegraphics[width=1\linewidth]{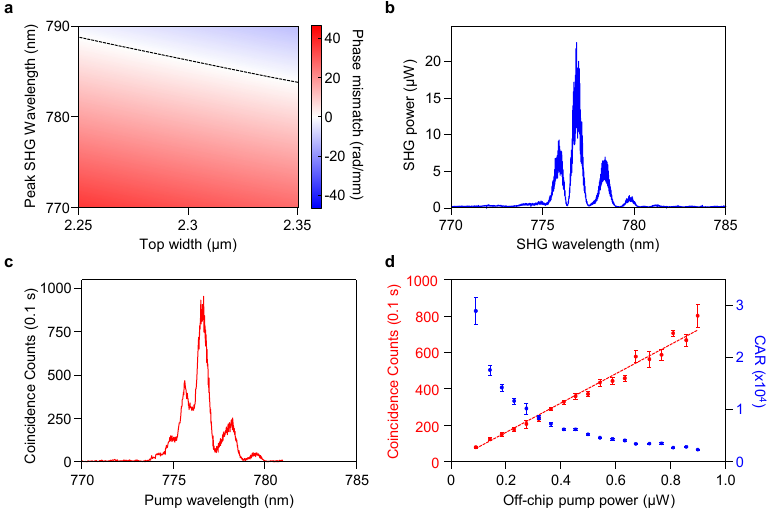}
    \caption{\label{fig:fig2} \textbf{Investigation on the single PPLN waveguide.} \textbf{a} Numerical calculation of the phase-matching map for second harmonic generation. The horizontal axis is the top width of the straight waveguide, and the vertical axis is the peak SHG wavelength. The black dashed line indicates the part where the phases are matched. \textbf{b} Experimentally measured second harmonic spectrum. \textbf{c} Experimentally measured coincidence counts as a function of pump wavelength. \textbf{d} Experimentally measured coincidence counts as a function of off-chip pump power. The error bar is standard deviation of five measurements. The red-dotted line is a linear fitting of the coincidence counts, where R-square value is 0.978. In \textbf{c} and \textbf{d}, the coincidence counts are measured with 100 ms exposure time and coincidence window is set as 2 ns.}
\end{figure}

To thoroughly investigate the poling period required for the QPM condition, we calculate the effective refractive index over a broad range of top widths, centered around the PSR top width of 2,304 nm, with values ranging from 2,250 nm to 2,350 nm. Figure~\ref{fig:fig2}a presents the phase mismatch map for varying waveguide top-width and fundamental/second harmonic (FH, SH) wavelengths when the poling period is set to 3.2 $\mu$m. This map indicates that SH light with wavelength between 783 nm and 788 nm can be generated with waveguide top-width ranging from 2,250 nm to 2,350 nm. A reference straight PPLN waveguide with a top-width of 2,304 nm shows a distinct SHG peak at 777 nm, confirming QPM with a 3.2 $\mu$m poling period, as shown in Figure~\ref{fig:fig2}b. The discrepancy between the numerically predicted and experimentally measured SH wavelength arises from marginal errors and inhomogeneities in the poling period, as well as variations in device geometries such as height, etch-depth, and sidewall angle. The detailed results of the impact of theses factors on the phase-matching condition is provided in the Supplementary Figure S2.

Next, we further investigate the photon pair generation and their brightness of PPLN device. The experimental setup for measuring coincidence counts utilizes fiber-based photon counting setup~\cite{kim2025freeform}. Figure~\ref{fig:fig2}c shows the coincidence count rates as a function of the pump wavelength, which nearly matches to the SHG spectrum measurements. To quantify the brightness of photon pair generation, we measured the coincidence count rates as a function of the pump power, as shown in Figure~\ref{fig:fig2}d. The measurement are repeated five times at each pump power level with exposure time of 100 ms. The fitted line clearly indicates a strong linear correlation between pump power and coincidence counts. From these results, the off-chip SPDC brightness is determined to be 8.059 MHz/mW. Taking into account the minimum measured coupling losses at the waveguide facets, approximately $-$4 dB for 1550 nm and $-$10 dB for 775 nm, the estimated on-chip brightness of photon pair generation reaches 508.5 MHz/mW. Also, the coincidence-accidental-ratio is calculated $2.9 \times 10^4$ at the power of 89.7 nW. These results demonstrate that our PPLN device exhibits significantly higher brightness compared to other integrated photonics material platforms that demonstrate the generation of the polarization Bell state~\cite{matsuda2012monolithically, wen2023polarization, zhang2024polarization, miloshevsky2024cmos}.

The PSR is a device that routes guided light based on its input polarization state~\cite{dai2011novel, wang2021efficient, chen2021broadband}. As briefly described in the previous section, the fundamental TE mode passes directly through the through port, while the fundamental TM mode undergoes polarization rotation and emerges as a fundamental TE mode through the cross port. More specifically, the PSR consists of two key regions: a rotation region, where the polarization of the mode is rotated, and a split region, where the output is split into the through and cross ports (see Supplementary Figure S3). The rotation of the mode is achieved through a mode-crossing phenomenon. In the rotation region, mode-crossing occurs between the fundamental TM mode and the $\mathrm{TE}_{10}$ mode. Unlike the fundamental TE mode, which maintains its polarization characteristics through the rotation region, the fundamental TM mode is converted into the TE$_{10}$ mode. The effective index of the TE$_{10}$ mode in the through branch matches that of the fundamental TE mode in the cross branch. As a result, the fundamental TM mode, after passing through the rotation region and being converted into the TE$_{10}$ mode, couples into the fundamental TE mode at the cross port. The determination of top-widths of the rotation and split regions are further discussed in Supplementary Note3. For generation of polarization entangled state, we use the PSR to translate the two distinct spatial modes into a polarization mode, where the two inputs are inserted into the through and cross ports, and output is the common polarization port.

\clearpage

\subsection*{Second harmonic generation of two PPLN waveguides}
\begin{figure}
    \includegraphics[width=1\linewidth]{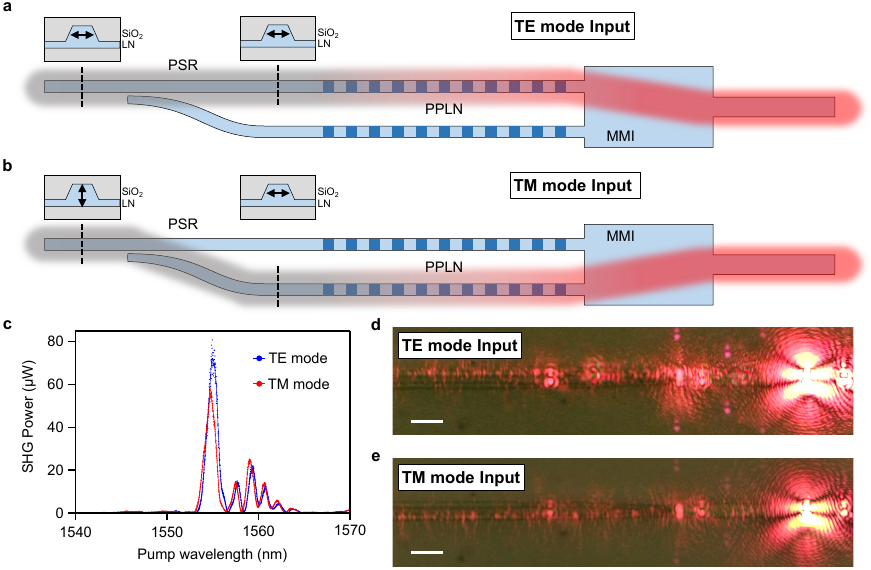}
    \caption{\label{fig:fig3} \textbf{Experimental results of second harmonic generation with PSR and two PPLNs.} \textbf{a,b} Conceptual illustration of second harmonic generation from telecom pump TE and TM mode input. \textbf{a} The TE mode input passes through the PSR in TE mode and meets the PPLN at the top of the figure. \textbf{b} The TM mode input changes path as it meets PSR and is converted to TE mode, meeting the PPLN at the bottom of the figure. \textbf{c} Experimentally measured SHG spectra for TE (blue) and TM (red) mode input. \textbf{d,e} Optical microscope image of second harmonic generation for (\textbf{d}) TE input pump and (\textbf{e}) TM input pump. Scale bars: 15 $\mu$m.}
\end{figure}

Subsequently, we carry out SHG measurements for both the through and cross ports of our PSR-PPLN device, as illustrated in Figure~\ref{fig:fig3}a and~\ref{fig:fig3}b. The objective of this experiment is to investigate how the path of SHG in PSR-PPLN waveguide is influenced by the polarization of the input pump light, as well as to compare the SHG spectra across the different paths. To this end, we couple a telecom band pump source into the output port of the PSR-PPLN waveguide using a lensed fiber, following a setup similar to that used for previous PSR measurements. The generated SHG signals are subsequently collected via another lensed fiber. To ensure adequate pump power, the input pump is amplified using an erbium-doped fiber amplifier (EDFA) prior to coupling it into the waveguide. A polarization controller (PC) is positioned after the EDFA to precisely adjust the polarization of the light for each path. Finally, the SHG characteristics are systematically measured while varying the input polarization between TE and TM modes.

Figure~\ref{fig:fig3}c presents the measured SHG spectra for different pump polarizations. For the through path, a peak wavelength of 1555.04 nm is observed when the pump is in the TE mode. Conversely, for the cross path, a peak wavelength of 1554.78 nm is observed under the TM mode pump input. Both peak wavelengths exhibit excellent agreement with the SHG peak wavelength of the reference waveguide in Figure~\ref{fig:fig2}b. The slight discrepancy between the peak wavelengths is likely due to minor fabrication imperfections, such as marginal errors in the poling process, and intrinsic thickness variations in the TFLN chip. The SHG spectrum corresponding to the TE mode pump demonstrates higher conversion efficiency. This can be explained by the additional coupling loss associated with the transition from input waveguide to cross port and the higher propagation loss for the TM mode.

\clearpage

\subsection*{Quantum state tomography of on-chip polarization entangled states}
\begin{figure}
    \includegraphics[width=0.95\linewidth]{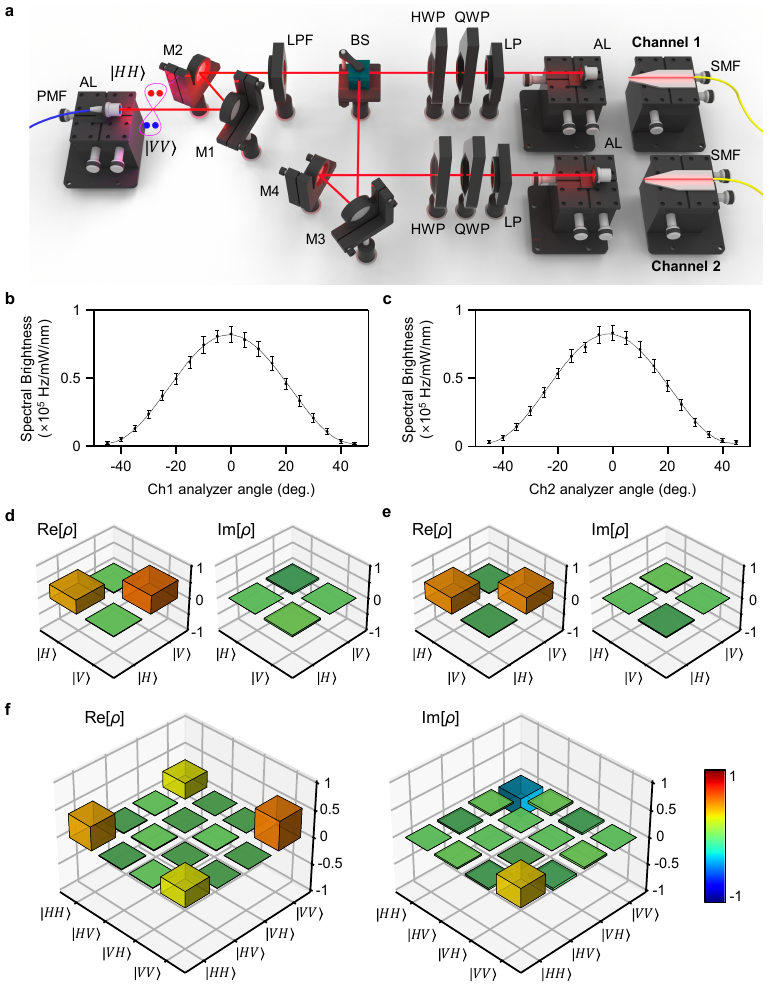}
    \caption{\label{fig:fig4} \textbf{Experimental results of quantum state tomography.} \textbf{a} Schematic illustration of experiment setup for quantum state tomography. PMF: Polarization maintaining fiber, AL: Aspheric lens, M: Mirror, LPF: Long pass filter, BS: Beam splitter, HWP: Half wave plate, QWP: Quarter wave plate, LP: Linear polarizer, SMF: Single mode fiber. \textbf{b,c} On-chip spectral brightness as a function of an analyzer (HWP) angle for \textbf{b} Channel 1 and \textbf{c} Channel 2. \textbf{d,e} Experimental result of single photon quantum state tomography for \textbf{d} Channel 1 and \textbf{e} Channel 2. \textbf{f} Result of two photon quantum state tomography.}
\end{figure}

Next, we perform the quantum state tomography (QST) to fully characterize the polarization states of photons generated on the TFLN chip. The experimental setup for QST utilizes a free space arrangement, as shown in Figure~\ref{fig:fig4}a. To preserve the polarization state of photons emitted from the waveguide during transfer to the QST setup, we used a polarization-maintaining lensed fiber. The photon pairs are coupled out to free space, and residual pump photons are removed by edge pass filters. Since the photon pairs occupy the same spatial mode, a 50:50 non-polarizing beam splitter is utilized for splitting the photon pair in to two distinct direction.

The polarization states of the photons are projected by employing a quarter-wave plate (QWP) and a half-wave plate (HWP) that mounted on the motorized states, followed by a vertically polarized fixed linear polarizer. Then, projected photons are coupled into the single-mode fibers, and their spectra are filtered using telecom dense wavelength division multiplexer (DWDM) to minimize of waveplate-induced errors. Photon counting is conducted using the superconducting nanowire single-photon detectors (SNSPDs). Polarization projection measurements are conducted across horizontal ($H$), vertical ($V$), diagonal ($D$), anti-diagonal ($A$), right-handed circular ($R$), and left-handed circular ($L$) bases.

Figure~\ref{fig:fig4}b and c shows the experimentally measured on-chip spectral brightness as functions of analyzer angles (rotation angles of the half wave plates) in channels 1 and 2, respectively. During each measurement, the analyzer angle of the other channel is fixed to vertical polarization. Measurements are repeated 30 times at each rotation angle with an exposure time of 100 ms, and error bars represent the standard deviation from these repeated measurements. Accidental coincidence counts are subtracted based on single-channel counts and a 200 ps coincidence window. The corresponding sinusoidal fittings, based on the theoretically predicted spectral brightness $P(\theta) = A \mathrm{cos}^2(2\theta-\theta_0) + B$ are included for comparison. Here, $A$, $B$, and $\theta_0$ represent the brightness amplitude, baseline offset, and measurement basis offset, respectively. In both panel, the experimentally measured on-chip spectral brightness exhibit good agreement with the sinusoidal fits and oscillate with high contrast, where calculated visibilities are $98.33 \%\pm 0.77\%$ and $97.36 \%\pm 0.82\%$. This shows that the the two photons are strongly entangled in their polarization states.

We further investigate the quantum state through density matrix reconstruction by QST, based on measured single-photon counts and coincidence counts across 36 bases~\cite{nielsen2010quantum, james2001measurement}. An exposure time of 500 ms and a coincidence window of 200 ps are used, with accidental counts similarly subtracted. The QST measurements were repeated 30 times to determine average values and uncertainties for the quality metrics.
Figure~\ref{fig:fig4}d and e show the reconstructed density matrices for the single photon states in channels 1 and 2, respectively. The calculated purities are $0.514$ and $0.510$, indicating that the both single photon states are mixed states.

Figure~\ref{fig:fig4}f presents the density matrix of the two photon state by the reconstruction results from the two photon QST, showing the high purity of $0.901 \pm 0.012$. Additionally, the concurrence of $0.900\pm 0.012$ confirms that the photon pairs are entangled in their polarization. The fidelity, calculated as $0.944 \pm 0.007$, takes into account efficiency difference between $H$ and $V$ polarization states observed in the single photon tomography and local phase change determined by the maximization of the fidelity. The observed slight discrepancies in the $H$ and $V$ polarization states are caused from the several practical factors, including nonlinear efficiency of the PPLN waveguides, the coupling efficiency of the PSR, the propagation loss difference between TE and TM modes in the waveguides, and the difference in coupling loss between the facet and lensed PM fiber of TE and TM modes. Despite these minor experimental challenges, the polarization entangled photon pairs are conclusively demonstrated. Additionally, the normalized ratio $\eta = 0.444$, and relative phase difference $2\phi = 0.867$ rad are calculated for fully describing the generated states.

\clearpage

\section*{Discussion and outlook}
In this study, we have successfully demonstrated a compact, high-quality, and bright source of polarization-entangled Bell state photons based on TFLN chip. Through the integrated combination of an MMI, two independent PPLNs, and a PSR, we experimentally achieved polarization-entangled photon pairs exhibiting exceptional quantum state characteristics. Notably, our device operates with a single continuous-wave (CW) laser at a low power level, yet exhibits strong performance metrics across all indicators of entanglement quality, including a concurrence of 0.900 and a fidelity of 0.944, along with a high purity of 0.901. Additionally, our PPLN achieves high brightness, with on-chip photon pair generation rate of 508.5 MHz/mW, significantly surpassing the silicon-based integrated photonics platforms.

Compared to previously reported TFLN-based time-bin entangled sources that employ a single PPLN waveguide and rely on a pulsed pump laser~\cite{finco2024time}, our approach not only simplifies the pump laser requirements but also demonstrates superior quantum performance in terms of concurrence and fidelity. Remarkably, this level of performance is achieved despite using two distinct PPLN waveguides, which inherently pose more stringent requirements on indistinguishability. This result not only highlights the intrinsic simplicity and robustness of polarization entanglement, but also indirectly indicates the scalability of our architecture.

Furthermore, when evaluated against on-chip silicon-based Bell state sources relying on SFWM, which inspired our device configuration, our TFLN-based design offers brighter SPDC photon pair generation, degenerate photon generation from a single pump, and a larger spectral separation between the pump and generated photons~\cite{miloshevsky2024cmos,matsuda2012monolithically}. In addition, the strong electro-optic capabilities of the TFLN platform pave the way for dynamic modulation the generated quantum states and implementation of multiplexing techniques~\cite{lin2022high,kaneda2019high}.

Our current device configuration offers promising pathways toward achieving even higher quality entanglement through further integration of components. In particular, the implementation of variable optical attenuation (VOA) could allow fine-tuning of photon pair generation efficiencies for TE and TM polarization paths, consequently improving the concurrence and purity of generated states~\cite{finco2024time}. Furthermore, the integration of electro-optic or thermo-optic phase shifter would facilitate accurate control and fine adjustment of the relative phase difference between TE and TM photon pairs. Incorporating active feedback and compensation components, including VOA and phase shifters can be promising candidate for improving the device functionality in aspect of device stability and operational robustness. Electro-optically tunable waveguide phase controllers, adaptive polarization tracking, and real-time quantum state monitoring would enable dynamic correction of polarization drifts and fluctuations induced by environmental disturbances~\cite{lin2022high}.

Despite the successful demonstration, experimental result also suggest room for further performance improvements, specifically concerning polarization crosstalk and extinction ratio. Residual polarization crosstalk, mainly caused from the fabrication imperfection in waveguide geometry especially of PSR. This can be identified as a limiting factor for achieving even higher concurrence and purity values.

Looking forward, the TFLN integrated photonic platform presents compelling potential beyond polarization entanglement, extending towards multidimensional quantum state encoding and hybrid entanglement schemes. Exploiting the TFLN platform, future developments might target quantum state engineering involving combination dimension of polarization, spatial, temporal, frequency-bin. By leveraging these higher-dimensional encoding strategies, researchers can unlock exponential scalability and versatility in quantum systems, significantly advancing quantum communications, high-capacity quantum networking, and multifunctional quantum sensing. Moreover, exploring non-degenerated frequency case could further extend the applicability of our platform in wavelength-multiplexed quantum network ~\cite{miloshevsky2024cmos}. In conclusion, our work emphasize the foundational components for scalable integrated quantum photonics. We expect our device can be regarded as promising platform in quantum technologies, driving advancements towards practical quantum information processing and applications.

\clearpage

\section*{Methods}
\subsection*{Device design and fabrication}
The effective mode indices are simulated using the finite element method with commercially available software, COMSOL Multiphysics. The design of PSR is performed with eigenmode expansion using commercially available software, Ansys Lumerical MODE.

Device fabrication involves two main steps, starting with inverting ferroelectric domains prior to waveguide fabrication. After cleaning the TFLN with 300 nm thickness, a Poly(methyl methacrylate) 950 A5 is deposited via spin coating at 2000 RPM. An additional e-spacer layer is spin-coated on the top of the resist to alleviate the charging effect during E-beam lithography. Subsequently, the poling electrode pattern is written using E-beam lithography (JEOL, JBX-9300FS). After development using a mixture of deionized water and isopropyl alcohol, a 100 nm thick chromium (Cr) layer is deposited via E-beam evaporator and lifted off using an acetone. To create the periodic domain inversion for quasi-phase matching, high-voltage electrical pulses are applied across the patterned electrode. Following periodic poling, the Cr electrodes are removed using chromium etchant, except for align markers. The schematic flow of periodic poling process is depicted in Supplementary Figure S4.

The next fabrication step is waveguide patterning. Hydrogen silsesquioxane (HSQ) resist is deposited using the spin coating at 2000 RPM, and an e-spacer layer is similarly applied. The waveguide patterns are defined using the same E-beam lithography system. After developing the HSQ resist using the AZ300MIF developer, inductively coupled plasma-reactive ion etching (ICP-RIE) with Ar ions is exploited to etch the waveguide pattern. Next, BOE is used to remove the remaining resist, and KOH solution is used for cleaning. SiO$_2$ cladding is subsequently deposited by a plasma-enhanced chemical vapor deposition. Finally, annealing process is carried out. The overall fabrication process is displayed in Supplementary Figure S5.

\subsection*{Device measurements}
For the second harmonic generation experiments, a tunable continuous-wave telecom band laser (Santec, TSL-710) serves as a pump laser, and its output is amplified by an erbium-doped fiber amplifier (Pritel). The polarization of the input pump is controlled using a polarization maintaining fiber-based linear polarizer (Thorlabs, ILP1550PM-APC) and a polarization-maintaining fiber bench equipped with a half-wave plate (Thorlabs, FBR-AH3) to achieve the desired linear polarization rotation. The transverse electric (TE) and transverse magnetic (TM) polarizations are applied using the polarization controller. The polarization controller ensures optimal polarization for efficient nonlinear interactions. Polarization-maintaining lensed fibers (OZ Optics), operating at a wavelength of 1550 nm and mounted on fiber rotator (Thorlabs, HFR007), are used to couple light into the TFLN waveguide. The birefringent axis of the polarization-maintaining fiber is aligned using a free-space linear polarizer and a reference 1550 nm laser. In contrast, single mode lensed fibers (OZ Optics), operating at a wavelength of 1550 nm are used to couple light out of the TFLN waveguides. We have measured the SHG signal intensity using a Si photodetector.

For quantum state tomography, a tunable near-infrared CW laser (New Focus, TLB-6712) is used for a pump source. The pump power is adjusted using a fiber-based variable optical attenuator (Thorlabs, VOA780-APC). Polarization management is performed using a fiber bench and fiber bench polarization optics (Thorlabs, FBR-AQ2, FBR-AH2). A single mode lensed fiber (OZ optics), operating at a wavelength of 775 nm, is utilized for the couples the pump source into the chip, while the output light is coupled to the polarization maintaining lensed fiber (OZ optics). Generated photons are collected and coupled out to the free space using the aspheric lens (Thorlabs, A280TM-C). Three cascaded long-pass filters (Thorlabs, FELH1050) are utilized for filtering out residual pump light. Photon pairs are splitted using a 50:50 non-polarizing beam splitter (Thorlabs, BS014). Polarization state projections are performed using half wave plates (Thorlabs, WPH05M-1550), quarter wave plates (Thorlabs, WPQ05M-1550) mounted on motorized rotation stage (Thorlabs, PRM1/MZ8), and linear polarizers (Thorlabs, LPNIRF050-MP2). The projected photons are coupled into single mode fibers using the aspheric lenses (Thorlabs, A280TM-C). Next, the dense wavelength division multiplexer (OZ optics) are utilized to filter the spectrum of photons. The photons are transferred to the superconducting nano-wire single photon detector with polarization controller. Detection signals are recorded by a time-correlated single-photon counting module (Qutools, QuTag-MC). The density matrix is reconstructed from measured data using the maximum likelihood estimation method.

\section*{Acknowledgments}
This work was supported by National Research Foundation (NRF) (2023M3K5A1094805,RS-2024-00343768,RS-2024-00509800), National Research Council of Science and Technology (NST) (CAP21034-000), and Korea Institute of Science and Technology (KIST) research program (2E33541,2E33571).

\section*{Conflict of interests}
\noindent The authors declare no competing financial interests.

\section*{Author contributions}
\noindent C.K, Hansol Kim, and Joonhyung Lee designed devices and experiments. C.K., M.C, S.M, and Jinil Lee fabricated the devices with help from H.H. and M.S.. C.K. and Hansol Kim performed measurements and analyzed data with help from Yong-su Kim. Hyounghan Kwon conceived the project with input from Yoon-ho Kim. Hyounghan Kwon and H.J. supervised the project. C.K., Hansol Kim, Hyounghan Kwon, and H.J. wrote the manuscript with help from Yong-su Kim. All authors discussed the results and commented on the manuscript.

\section*{Data Availability Statement}
\noindent The data that support the findings of this study are available from the corresponding author upon request.

\clearpage
\section{References}
\bibliographystyle{naturemag_noURL}
\bibliography{reference}
\clearpage

\clearpage

\clearpage



\clearpage

\end{document}